# P-dit Probabilistic Ising Machine for Solving the Quadratic Assignment Problem

Christian Duffee[1,†], Chadbourne M. Burling-Smith[1,†], Jordan Athas[1], Andrea Grimaldi[2], Giovanni Finocchio[3], Ermin Wei[1,4], and Pedram Khalili Amiri[1,5,*]

**[1]Department of Electrical and Computer Engineering, Northwestern University, Evanston, IL, United States**

**[2]Department of Electrical and Information Engineering, Politecnico di Bari, Bari, Italy**

**[3]Department of Mathematical and Computer Sciences, Physical Sciences and Earth Sciences, University of Messina, Messina, Italy**

**[4]Department of Industrial Engineering and Management Sciences, Northwestern University, Evanston, IL, United States**

**[5]Applied Physics Program, Northwestern University, Evanston, IL, United States**

**[*]Corresponding author: pedram@northwestern.edu**

**†These authors contributed equally to this work**

## Abstract

Combinatorial optimization problems represent a wide range of real-world scenarios where complicated interactions make it difficult to find the best solution. One example is the quadratic assignment problem (QAP), which involves determining the optimal placement of facilities at set locations which minimizes the products of material flow and facility distance. This representation is descriptive of many real-world scenarios, including the aggregate transportation costs of a supply chain. In this work, a probabilistic Ising machine (PIM) approach is implemented using probabilistic d-dimensional variables (p-dits), which are generalized, multi-state and multi-dimensional extensions to probabilistic bits (p-bits). Each p-dit corresponds to a location and stochastically oscillates between facility assignments based on the influence of the other p-dits. We show that with the same runtime and CPU, the PIM finds the best-known solution on 95% of considered instances from the QAP Library dataset, compared to just 36% for the standard Gurobi solver. For the unique largest problem in the library, a 2 to 3 order-of-magnitude decrease is observed in the time needed to reach specific solution qualities. We also show parallelization of our PIM through GPU implementations. A comparison to state-of-the-art QAP solver algorithms shows that they are consistently outperformed by both CPU and GPU implementations of the p-dit Ising machine.

## Introduction

Combinatorial optimization encompasses a broad class of problems which involve finding some ordered, or unordered, combination of elements from a set. Combinations must meet a list of problem-specific constraints to be valid solutions, with the quality of each solution judged based on a goal function. Combinatorial optimization problems (COPs) are often non-deterministic polynomial-time hard, becoming rapidly more difficult to solve as their size increases. Nevertheless, finding high-quality solutions is vital for many logistical[1], resource sharing[2-4], routing[5,6], and financial[7] applications.

A particularly important COP is the quadratic assignment problem (QAP), which can be used for a variety of applications such as cost minimization in supply chains, warehouse management, and process communication[8]. The QAP involves finding, in symmetrical cases,

$$\min_{p} \sum_{i=1}^{N} \sum_{j=1}^{N} D_{ij} F_{p(i)p(j)} \tag{1}$$

where $p(i)$ maps to the index (1 to $N$) of the facility assigned to location $i$, and $D$ and $F$ are $N \times N$ sized matrices. These matrices are the distance matrix and flow matrix, respectively, and can be understood as referring to the distance between potential facility locations and the flow of goods between facilities that can be assigned to them. Each term of the double summation is then the product of the distance between location $i$ and location $j$ with the flow between the facilities currently assigned to the two locations. The solution to the problem involves finding the optimum pairing between facilities and facility locations, such that the sum of the flow-distance products is minimized[9-11].

In an unrestricted instance, there are $n!$ possible combinations and $n^4$ flow-distance product terms to consider. The large computational cost of evaluating candidate solutions can lead to relatively small problem instances (e.g., $n < 50$) remaining intractable on modern hardware[9]. As such, there is large room for improvement in algorithmic techniques to generate high-quality solutions for practical applications.

COPs can be transformed into a quadratic unconstrained binary optimization (QUBO)-like format, the Ising model, where an energy function is represented as the sum of the scaled product of two state ($-1$ and $+1$) spins. These spins can be updated in a continuous manner, such as in coherent Ising machines[12-16] and simulated bifurcation machines[17-19], or in a discrete-time manner, such as in probabilistic Ising machines (PIMs)[20,21]. While both can find minima of energy landscapes, which correspond to the solutions of the original COPs, the latter may prove easier to implement on digital hardware[22-24]. Their energy function can be decomposed into a series of interactions between two-state spins which collectively creates an effective influence force ($I$) on each spin, biasing them towards either the $-1$ or $+1$ state. These spins can be implemented as p-bits[21,24]: discrete-time oscillators that favour a state based on their $I$, scaled by an inverse mathematical temperature, $\beta$. As individual p-bits in a PIM are probabilistically updated, the system as a whole converges to a low-energy configuration. This process is effective in solving many disparate classes of COPs, including QAPs[25].

P-bits can be viewed as probabilistic variables which hold one of two states along a single dimension. An extension to this concept, shown in Fig. 1a, are probabilistic d-dimensional variables (p-dits), which can hold multiple states along multiple dimensions[26]. This allows a more natural representation of non-binary variables, leading to two primary advantages. The first is the elimination of the need for additional energy terms that map between the innate multi-dimensional structure of a problem and a QUBO formulation. The second, and more fundamental, is better exploration of the configuration space, which can result in improved performance[26-28]. In this work, we consider a specific restriction to this general definition, where each p-dit holds a state representing a unit vector along one of $d$ dimensions, as shown in Fig. 1b. The $I$ for each p-dit then becomes a vector, with each element exerting a force towards one of the dimensions. Whenever a p-dit is updated, its state can transition to any of these dimensions or remain in the current one.

In this work, we describe how the general QAP can be mapped into a p-dit-based PIM. We then report the performance of a central processing unit (CPU) implementation in solving problems within the Quadratic Assignment Problem Library (QAPLIB)[9] dataset, compared to traditional optimization software. We follow with reporting on the performance of the same p-dit PIM algorithm on a graphics processing unit (GPU). We also discuss a modification of this algorithm which implements concurrent updates and its performance when solving QAPLIB problems. We compare each of these three methods to prior reported results and conclude with a perspective on future research and development opportunities.

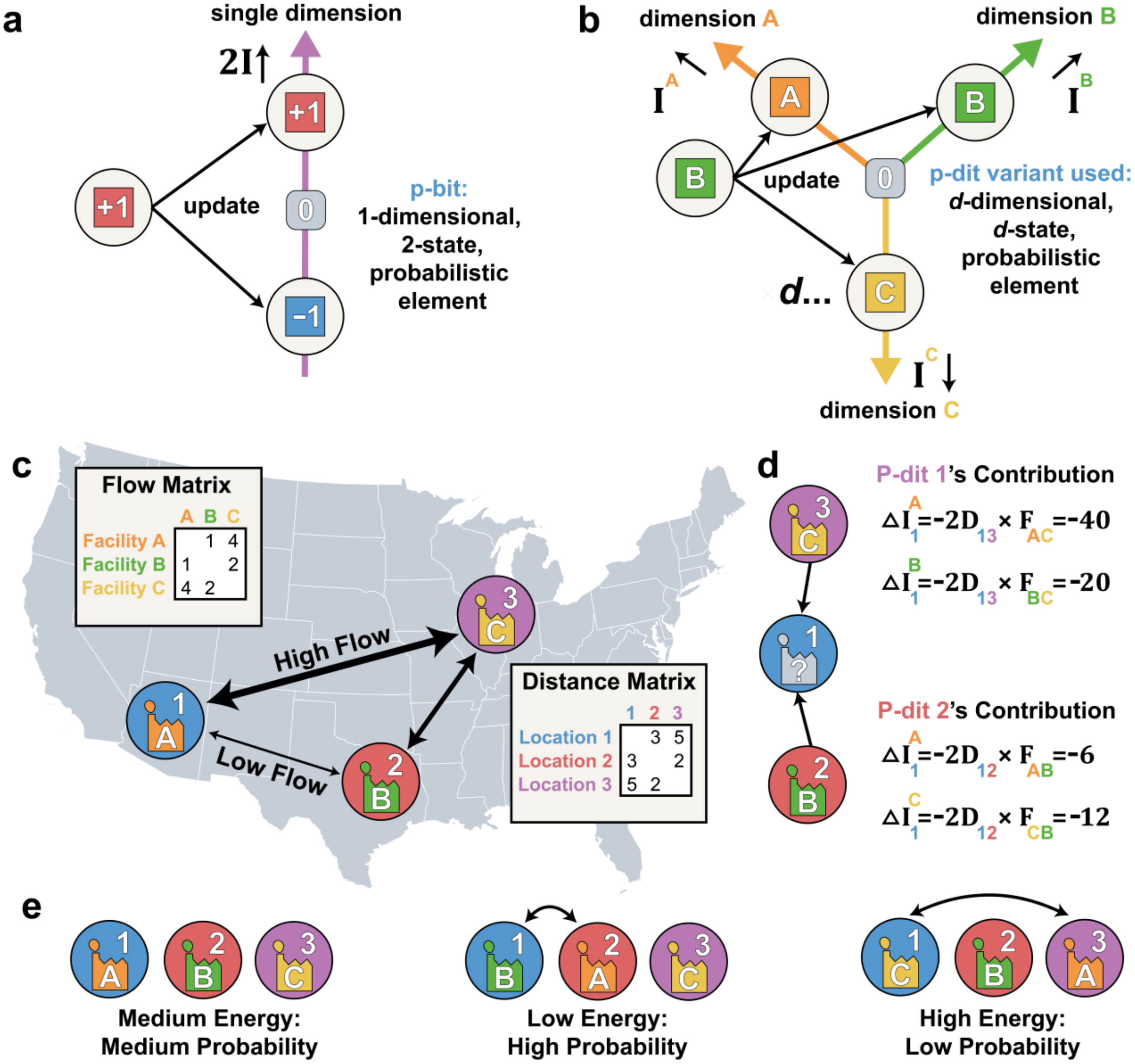

**Fig 1 | Overview of QAP problem and probabilistic calculations. a** The update scheme of a p-bit. When it is updated, a p-bit enters either the $+1$ or $-1$ state, which are opposites along a single dimension. A single input variable ($I$) pushes it towards (for positive values) or away from (for negative values) the +1 state. **b** The update scheme of the p-dit variant used in this work. When it is updated, the p-dit can align along any of $d$ dimensions. A $d$-element input vector biases the p-dit towards or away from each of these dimensions independently. **c** Graphical depiction of a three-facility instance of a QAP problem. Insets show the flow and distance matrices between the facilities and locations, respectively. Each location is represented by a p-dit, which holds a state that corresponds to a facility. **d** Depiction of the calculation of the influence force

vector for updating the p-dit which represents location 1. The other p-dits bias location 1's p-dit away from the facilities which have high flow with their own, scaled by the distance between the locations. **e** Depiction of the three possible new states produced by updating the p-dit for location 1: either no changes can be made, locations 1 and 2 have their facilities swapped, or locations 1 and 3 have their facilities swapped. Lower-energy states are more likely to occur.

# Results

## Implementation

The PIM for each QAP is composed of $N$ p-dits, each corresponding to a single location. Every p-dit holds a state of a unit vector along one of $N$ dimensions, each representing a facility. An input "force" is exerted along each dimension based on the state of all other p-dits. Each will attempt to push other p-dits away from facilities that have high flow rates with their current facility—an effect that scales with the distance between the two. If $I_i^a$ represents the force on the $i^{th}$ location towards the $a^{th}$ facility, then

$$I_i^a = -\sum_{j=0}^{N} D_{ij} F_{a,p(j)} + D_{ji} F_{p(j),a} \,. \tag{2}$$

Each p-dit is initialized with a unique dimension, and only swaps of facilities between p-dits are considered during updates. This prevents the occurrence of any invalid configurations, where any facility is assigned to more than one location, eliminating the need to introduce energy terms to enforce such a requirement. An example of the problem is illustrated in Fig. 1c-e.

Without loss of generality, the swapping of the $i^{th}$ p-dit which is along the $a^{th}$ dimension with the $j^{th}$ p-dit which is along the $b^{th}$ dimension is

$$\Delta E_{i\leftrightarrow j}^{a\leftrightarrow b} = -\left[I_i^b - I_i^a + I_j^a - I_j^b - \left(D_{ii} + D_{jj} - D_{ij} - D_{ji}\right)\left(F_{aa} + F_{bb} - F_{ab} - F_{ba}\right)\right]. \tag{3}$$

If D is symmetric and diagonal elements are 0, as in the physical case, instead we can use

$$\Delta E_{i\leftrightarrow j}^{a\leftrightarrow b} = -\left(I_i^b - I_i^a + I_j^a - I_j^b\right). \tag{4}$$

For each p-dit update, we consider the possibility of swapping with each of the other p-dits, as well as remaining in the current state. The update probability of each swap is then

$$P(i \leftrightarrow j) = \frac{\exp\left(-\beta \Delta E_{i\leftrightarrow j}^{p(i)\leftrightarrow p(j)}\right)}{\sum_{k=1}^{n} \exp\left(-\beta \Delta E_{i\leftrightarrow k}^{p(i)\leftrightarrow p(k)}\right)}. \tag{5}$$

During each iteration, every p-dit initiates an update once, leading to an iteration runtime that roughly grows with $N^2 \log(N)$.

Parallel tempering is used to help prevent the PIM from spending significant time holding configurations that correspond to a local minimum of the energy function[29]. Multiple replicas of the same system of p-dits are used with varying values of $\beta$. Replicas with higher temperatures broadly explore the energy landscape, while those with lower temperatures find the minimum within a narrow region. After a set number of update iterations, the system configurations are themselves swapped with the aim of passing lower energy configurations from high- to low-$\beta$ replicas, as depicted in Fig. 2a. Even with parallel tempering, it is possible for high-$\beta$ replicas to get stuck in local minima and not give the configurations found by more random replicas a chance. To help minimize this, we keep track of the time since the highest-$\beta$ replica has been swapped; if this exceeds a threshold, it is then swapped with the lowest-$\beta$ replica.

To select the range of $\beta$ values used, a quick sweep procedure was used. In it, a series of short trials, without swaps, at different $\beta$ values were performed. When the ending energies of each value were averaged, they formed a sigmoidal-like shape with an uptick at the end representing frozen conditions, as shown in Fig. 2b. The lowest $\beta$ was taken as that which achieved ~40% of the difference between the minimum and maximum energy found. The highest $\beta$ value was two steps beyond that which achieved the minimum energy. This range was selected to maximize effective exploration of the energy landscape, as values below are too random, and those above are too close to pure gradient descent. To achieve parallelization, these sets of replicas can be duplicated, with each attempting to solve the problem independently. Further implementation details are included in the Methods section.

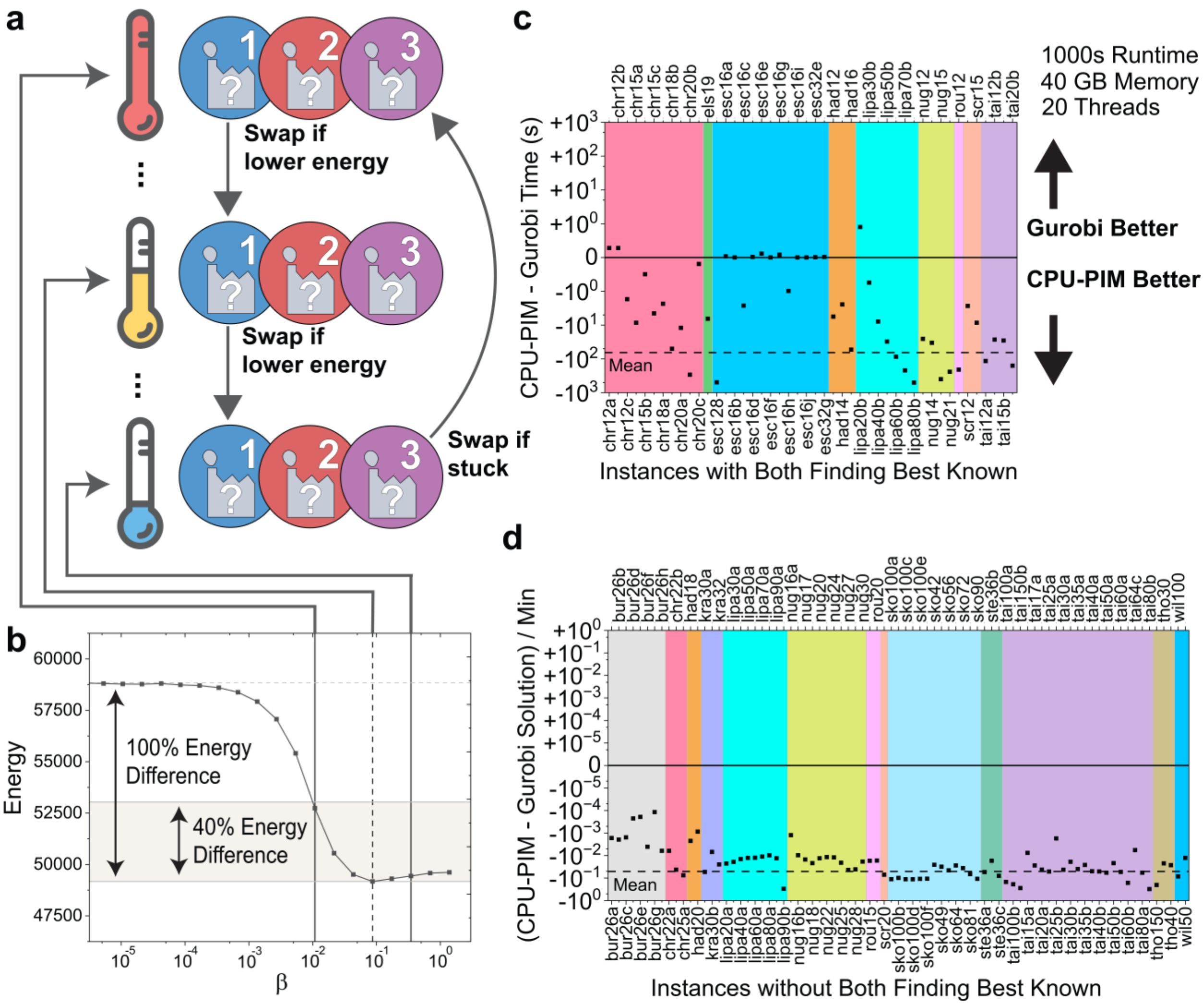

**Fig 2 | Parallel tempering and the relative performance of the CPU-PIM compared to Gurobi. a** A conceptual depiction of parallel tempering. Each replica runs at a different temperature, with configurations being regularly swapped from hotter, or more random, replicas to colder, or more deterministic, replicas if

their energy values are lower. If no configurations have been swapped into the coldest replica for a specific timeout, the coldest and hottest replica's configurations are swapped to speed convergence. **b** Data from a $\beta$ sweep. Short trials are run at increasing values of $\beta$ to characterize each problem. The $\beta$ for the hottest replica is taken as the point where the energy drops to around 40% between the minimum and maximum found. The $\beta$ for the coldest replica is taken as two sample points past where the minimum energy was found. **c** Runtimes for QAPLIB instances where both Gurobi and the CPU-PIM found the best known solution. Values below zero indicate that the CPU-PIM found the solution faster than Gurobi. **d** Solutions for QAPLIB instances where at least one of Gurobi and the CPU-PIM did not find the best known solution. Values below zero indicate the CPU-PIM found a higher-quality solution than Gurobi.

## Performance

Instances from the QAPLIB[9] were used to benchmark the performance of a CPU implementation of the described approach (CPU-PIM). The problems are labelled using three letters corresponding to the work they originate from, followed by their number of facilities, and finally, if multiple of the same size are included, a single indexing letter. Each problem was attempted once, using 20 independent CPU threads, each with 10 replicas. A memory limit of 40 GB and time limit of 1000 seconds for each problem was used. More details on simulation parameters are included in the Methods section.

A standard optimizer tool, Gurobi[30] was used as a baseline with the same thread count, memory limit, and time limit as with the PIM. The representation of QAP problems used for it is described in the Methods section. Gurobi's memory utilization tended to be much greater than the PIM's, and the 40 GB memory limit was reached on the two 150 facility problem instances. Gurobi was able to find the best known solution on 46 of the 127 instances attempted. The CPU-PIM, however, found the best known solution for 121 instances. More detailed comparisons between the performance of Gurobi and the PIM are shown in Fig. 2c-d.

For some problem instances, both solvers found the best known solution within the allotted time limit. For these, the time each solver took to first find the solution is compared in Fig. 2c. Negative values indicate that the CPU-PIM found the solution earlier, while positive values indicate the opposite. The two extreme differences are instances in which Gurobi and the PIM outperformed the other by 0.9 and 504.2 seconds, respectively. On average, the PIM outperformed Gurobi by 66.0 seconds, finding the solution faster in 81% of problem instances.

The relative solution qualities achieved on problem instances for which at least one of the solvers did not find the best known solution are shown in Fig. 2d. While on six instances neither solver found the best known solution, the CPU-PIM was the single solver which found it on the remainder of the instances. The difference in quality between the two solutions, divided by the minimum of the two, is graphed, where negative values show better performance by the CPU-PIM.

The largest problem, tai256c, was treated differently due to its symmetries by both Gurobi and the CPU-PIM. Implementation details are included in the Methods section. The evolution of the found solution over a long trial length is shown in Fig. 3 with both solvers under the same constraints. The Gurobi solver made steady improvement over the ~10.7-hour runtime but was unable to find the best known solution. Meanwhile, the first CPU-PIM thread, and thus the CPU-PIM as a whole, found the best known solution in ~517 seconds (8.6 minutes). By the end of the trial, 18 of the 20 PIM threads found the best known solution. Moreover, throughout nearly the entire runtime, every single independent CPU-PIM thread outperformed Gurobi. Overall, the CPU-PIM maintained a two to greater than three order of magnitude lead in the runtime required to reach each specific solution quality.

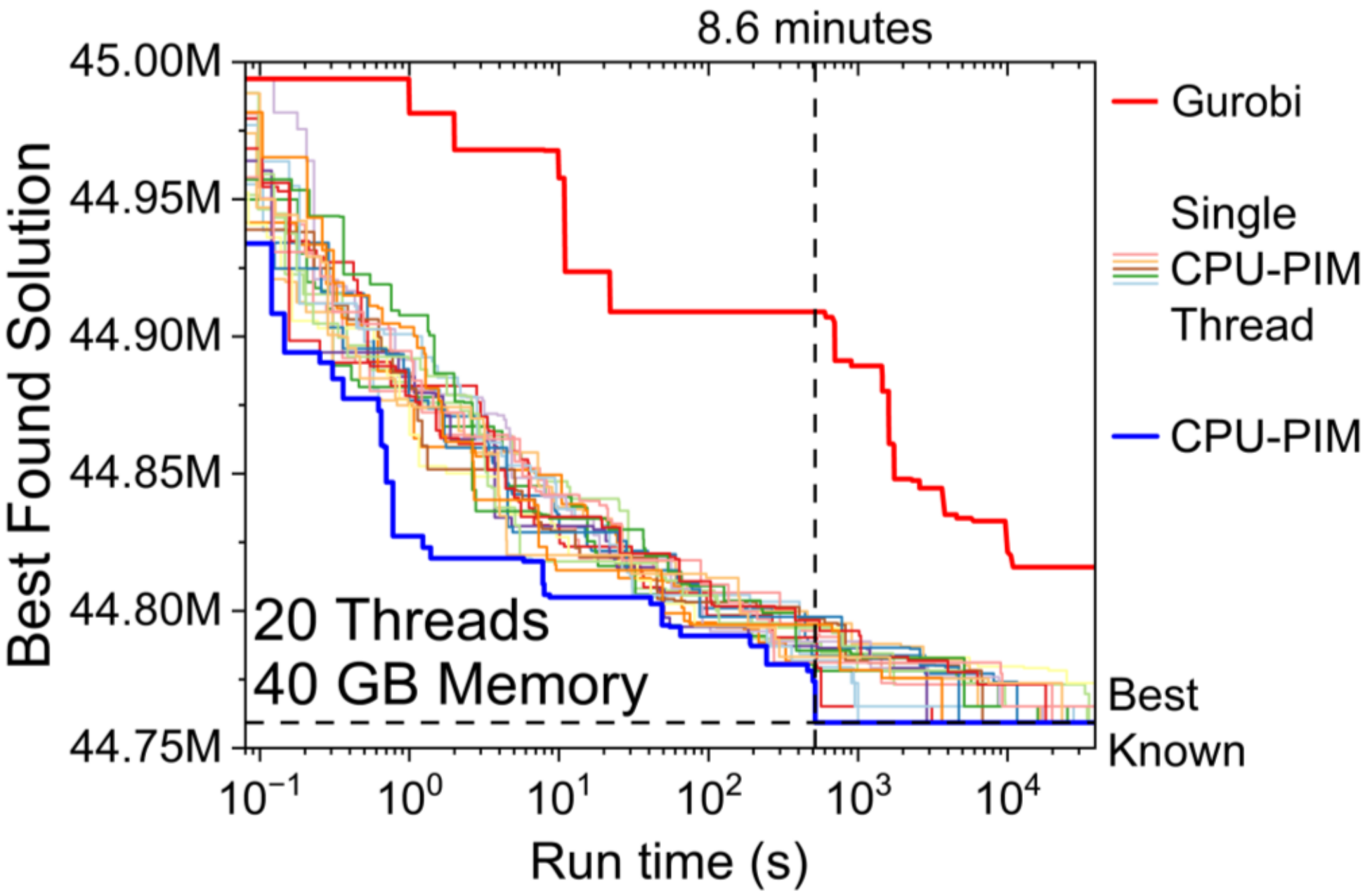

**Fig 3 | Time evolution of largest QAPLIB problem.** The best found solution by the Gurobi solver, each of the independent 20 PIM threads, and the PIM as a whole are plotted for the largest instance in the QAPLIB dataset. Over the course of the ~10.7-hour trial, Gurobi never found the best reported solution. Meanwhile, the best PIM thread found it in ~517 seconds (8.6 minutes). By the end of the trial, all but two of the 20 PIM threads had found the best known solution.

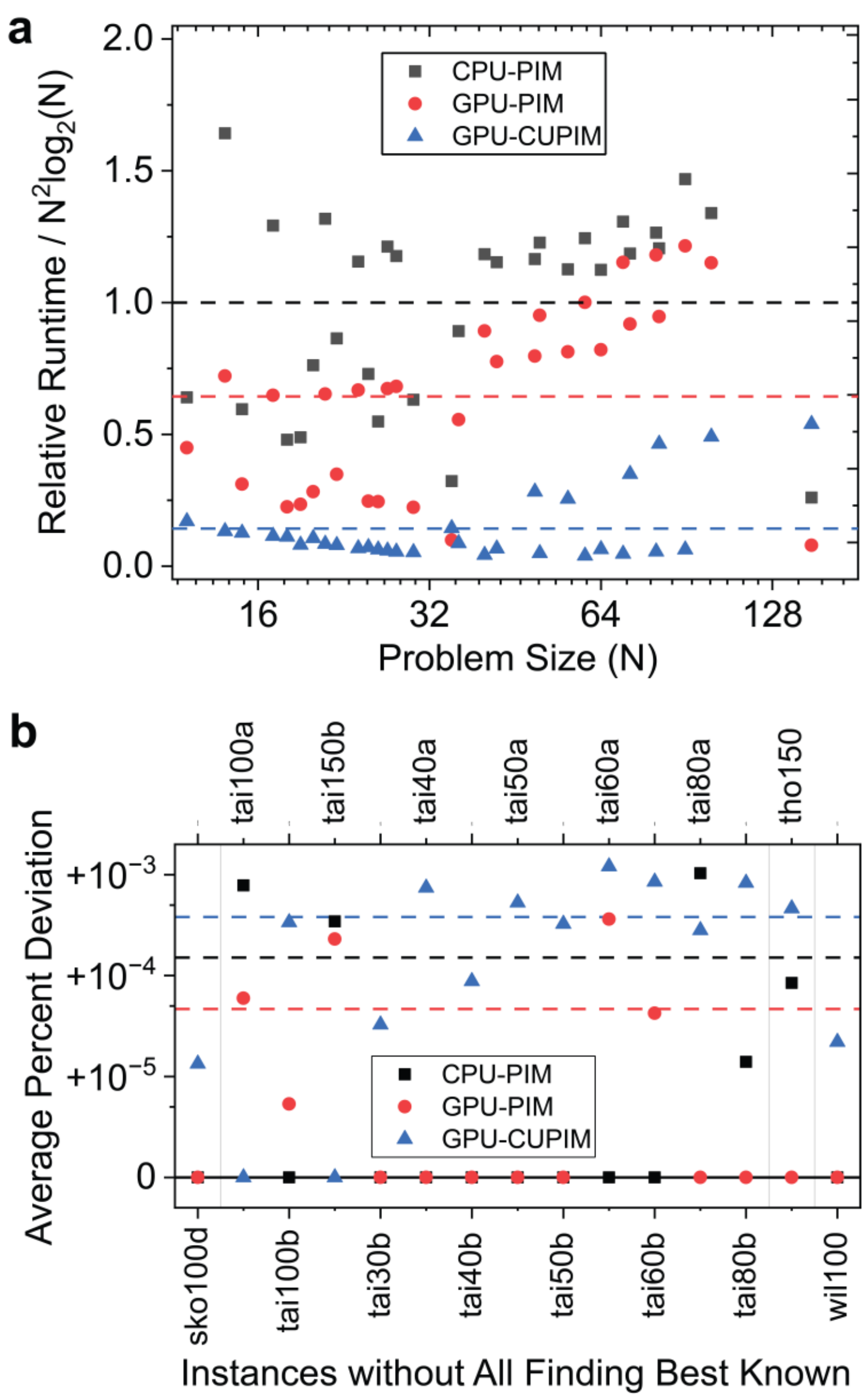

**Fig 4 | Relative performance of the GPU-PIM compared to the CPU-PIM on QAPLIB problems.** The relative performance between the GPU-PIM and the CPU-PIM implementation solver on all problems from the QAPLIB dataset. **a** Instances where one of the two solvers found the best reported solution; for 10 out of 14 problem instances this was the GPU-PIM solver. **b** Instances where neither solver found the best known solution; 12 out of 16 had the GPU-PIM finding the higher quality solution. **c** Instances where both solvers found the best reported solution; the GPU-PIM solver found it on average 20.34 seconds quicker than the CPU-PIM implementation.

## GPU Implementations

Due to their many computational cores, GPUs excel at performing massively parallel computations, however, they have relatively weak serial performance[31-33]. The CPU-PIM implementation achieves some parallelism by duplicating the same problem across several threads. For a GPU implementation (GPU-PIM) of the same algorithm, we expand this by parallelizing across both duplications and replicas at the block level. Unlike with the CPU-PIM, the number of problem duplicates is not tied to physical hyperthreading hardware, ensuring greater portability between different devices. Further parallelization is achieved during updates; both the calculation of the probability contributions of other p-dits and changes in influence force ($I$) post-swap are paralellized across threads.

Due to the presence of all-to-all p-dit coupling interactions, the updates themselves cannot be made in parallel without affecting the convergence dynamics of the system. That is, every swap that occurs modifies the probability of every subsequent possible swap. Nevertheless, the increased speed of parallel updates may justify the deviations from the ideal dynamics. Importantly, these are minimized at later iterations and for high values of $\beta$, as the configuration becomes stable and most updates result in a p-dit maintaining its current state.

We explored this approach with a GPU implementation of a concurrently updating PIM (GPU-CUPIM). The GPU-CUPIM performs the updates of an iteration in parallel by (1) computing the energy differences of all possible swaps using $I$, (2) resolving conflicts between overlapping swap proposals by prioritizing those that result in a larger energy decrease, (3) applying all accepted swaps simultaneously, and (4) updating the force and energy matrices.

The base speed of the three approaches is compared in Fig. 4a, across problems of different sizes, for a constant 40,000 iterations. The GPU-PIM is able to complete, on average, iterations in 61% of the time required by the CPU-PIM, while the GPU-CUPIM can do so in just 12% of the time. These differences do not directly translate into better solution quality, however, as demonstrated in Fig. 4b, on the problem for which the solution quality of the three solvers differ. While the GPU-PIM does indeed outperform the CPU-PIM in terms of average percent deviation (APD), the GPU-CUPIM produces lower quality solutions, due to its deviations from the ideal probabilistic dynamics. Nevertheless, it still found the best known solution in 88% of the total problems explored, compared to the 36% achieved by Gurobi.

## Comparison with Other QAP Solvers

| **Solver** | **CPTS**[34] | | **PHA**[35] | | **PABC-QAP**[36] | | **BLS**[37] | | **CPU-PIM** | | **GPU-PIM** | | **GPU-CUPIM** | |
|---|---|---|---|---|---|---|---|---|---|---|---|---|---|---|
| **Processors** | 10 | | 10 | | 255 | | **1** | | **1** | | **1 GPU** | | **1 GPU** | |
| **External Problem-Specific Parameters** | **No** | | **No** | | **No** | | Yes | | **No** | | **No** | | **No** | |
| | **APD** | **Time (min)** | **APD** | **Time (min)** | **APD** | **Time (min)** | **APD** | **Time (min)** | **APD** | **Time (min)** | **APD** | **Time (min)** | **APD** | **Time (min)** |
| tai20a | - | 0.1 | - | 0.4 | - | **0.0** | - | **0.0** | - | **0.0** | - | **0.0** | - | **0.0** |
| tai25a | - | 0.3 | - | 0.6 | - | 0.1 | - | **0.0** | - | **0.0** | - | **0.0** | - | **0.0** |
| tai30a | - | 1.6 | - | 1.0 | - | 0.2 | - | **0.0** | - | **0.0** | - | **0.0** | - | **0.0** |
| tai35a | - | 2.3 | - | 1.3 | - | **0.0** | - | 0.2 | - | 0.1 | - | **0.0** | - | **0.0** |
| tai40a | 0.148 | 3.5 | - | 10.6 | - | 4.0 | 0.022 | 38.9 | 0.007 | 3.6 | - | **1.3** | 0.074 | 0.7 |
| tai50a | 0.440 | 10.3 | - | 12.7 | 0.312 | 19.8 | 0.157 | 45.1 | - | 9.0 | 0.016 | **3.5** | 0.088 | 7.8 |
| tai60a | 0.476 | 26.4 | - | 19.6 | 0.449 | 34.0 | 0.251 | 47.9 | 0.103 | 10.6 | 0.106 | 8.1 | 0.150 | 7.3 |
| tai80a | 0.570 | 94.8 | 0.644 | 40.0 | 0.827 | 173.0 | 0.517 | 47.3 | 0.435 | 11.9 | **0.396** | 10.6 | 0.417 | 9.7 |
| tai100a | 0.558 | 261.2 | 0.537 | 71.9 | 0.644 | 335.6 | 0.430 | 39.0 | 0.405 | 12.0 | **0.326** | 12.1 | 0.361 | 9.3 |
| tai20b | - | 0.1 | - | 0.4 | - | **0.0** | - | **0.0** | - | **0.0** | - | **0.0** | - | **0.0** |
| tai25b | - | 0.4 | - | 0.6 | - | **0.0** | - | **0.0** | - | **0.0** | - | **0.0** | - | **0.0** |
| tai30b | - | 1.2 | - | 0.8 | - | 0.1 | - | **0.0** | - | **0.0** | - | **0.0** | 0.003 | **0.0** |
| tai35b | - | 2.4 | - | 1.1 | - | **0.0** | - | **0.0** | - | 0.1 | - | **0.0** | 0.017 | 4.1 |
| tai40b | - | 4.5 | - | 1.6 | - | 0.2 | - | **0.0** | - | 0.1 | - | **0.0** | 0.002 | **0.0** |
| tai50b | - | 13.8 | - | 5.8 | - | 1.6 | - | 2.8 | - | **0.1** | - | **0.1** | 0.032 | 7.5 |
| tai60b | - | 30.4 | - | 9.5 | - | **0.7** | - | 5.6 | 0.009 | 1.8 | 0.001 | 3.5 | 0.063 | 4.4 |
| tai80b | - | 110.9 | - | 27.7 | - | **8.0** | - | 11.4 | 0.005 | 6.4 | 0.014 | 4.9 | 0.044 | 6.9 |
| tai100b | 0.001 | 241.0 | - | **42.5** | - | 164.7 | - | 16.0 | 0.002 | 4.1 | 0.001 | 4.7 | 0.013 | 10.2 |
| sko42 | - | 5.3 | - | 1.6 | - | 0.2 | - | 1.7 | - | 0.1 | - | 0.1 | - | **0.0** |
| sko49 | - | 11.4 | - | 4.0 | - | 7.5 | - | 0.5 | - | 0.2 | - | **0.1** | - | **0.1** |
| sko56 | - | 21.0 | - | 16.2 | - | 0.9 | - | 1.1 | - | 0.2 | - | **0.1** | - | **0.1** |
| sko64 | - | 42.9 | - | 23.1 | - | 2.4 | - | 1.3 | - | 0.3 | - | 0.2 | - | **0.1** |
| sko72 | - | 69.6 | - | 33.6 | - | 7.0 | - | 4.1 | - | 0.6 | - | **0.3** | - | 2.2 |
| sko81 | - | 121.4 | - | 39.9 | - | 15.2 | - | 13.9 | - | 0.8 | - | **0.3** | - | 6.1 |
| sko90 | - | 193.7 | - | 40.5 | - | 14.7 | - | 16.6 | - | 1.1 | - | **0.5** | 0.002 | 5.2 |
| sko100a | - | 304.8 | - | 41.7 | 0.008 | 170.5 | 0.001 | 20.8 | - | 1.5 | - | **0.7** | 0.003 | 6.0 |
| sko100b | - | 309.6 | - | 42.3 | - | 127.7 | - | 10.8 | - | 1.3 | - | **0.6** | - | 3.9 |
| sko100c | - | 316.1 | - | 42.2 | - | 48.3 | - | 15.5 | - | 1.5 | - | **0.7** | - | 5.8 |
| sko100d | - | 309.8 | - | 41.9 | - | 45.9 | 0.001 | 38.9 | - | 2.7 | - | **1.2** | 0.003 | 6.6 |
| sko100e | - | 309.1 | - | 42.5 | - | 42.1 | - | 42.5 | - | 1.4 | - | **0.6** | 0.001 | 7.7 |
| sko100f | 0.003 | 310.3 | - | 42.0 | - | 28.3 | - | 17.3 | - | 1.4 | - | **0.6** | 0.002 | 3.6 |
| **Average** | 0.071 | 101.0 | 0.038 | 21.3 | 0.072 | 40.4 | 0.044 | 14.2 | 0.031 | 2.4 | **0.028** | **1.8** | 0.041 | 3.7 |

**Table 1| Comparison with state-of-the-art dedicated QAP algorithms.** A comparison between the three methods explored in this work and a series of dedicated QAP algorithms reported in the literature.

The CPU-PIM, along with the GPU-PIM and the GPU-CUPIM, was compared with several high-performing QAP algorithms reported in the literature on 31 of the problems in QAPLIB. These results are shown in Table 1. As the hardware is different, it is difficult to compare results directly. Nevertheless, both the CPU-PIM and the GPU-PIM consistently achieved better solutions, compared to each of the established solvers. Of these, only the parallel hybrid algorithm[35] (PHA) achieved lower scores than the GPU-CUPIM. All three of the PIM implementations achieved average runtimes faster than each of the comparisons.

## Discussion

We have shown how p-dit probabilistic Ising machines can be constructed to solve QAP instances. With parallel tempering, a PIM can find solutions of much higher quality compared to the state-of-the-art Gurobi solver on problems from the standard QAPLIB. While limited parallelization is achieved with a CPU implementation, by running multiple PIM instances on separate threads, greater parallelization is needed for a GPU implementation. This can be done both across replicas and other p-dits for performing update calculations. Additional parallelization across updates themselves can be achieved at the expense of breaking the mathematical convergence dynamics.

Future work is needed to expand these methods to other similar problems of interest, such as traveling salesman variants, where p-dit dimensions can represent some unique variable assignments that can be swapped. We expect that similar improvements in performance are to be found in the basic PIM case, however, the extent to which interfering p-dits can be updated simultaneously in a highly parallelized implementation will need to be determined on a problem-by-problem basis.

Beyond CPU and GPU hardware implementations, field-programmable gate arrays and application-specific integrated circuit p-dit PIMs should be explored. While usually more expensive, these platforms allow more flexibility in the implemented architecture which can be tailored to the nature of the problem, thus potentially leading to greatly improved performance. Specifically, with application-specific integrated circuits, high-speed and high-density hardware entropy sources, such as magnetic tunnel junctions[20,24,25,38], can be directly integrated to result in further performance enhancement.

# Methods

## CPU-PIM Algorithm

The PIM's algorithm started with an optional $\beta$-range selection period. This was used for the data presented in Fig. 2 and Table 1. For a wide range of $\beta$ values, the ending energies (rather than the best-found energies) after 400 iterations were averaged. This portion of the algorithm was generally fast relative to the primary portion.

During the main portion of the algorithm, 20 completely independent duplications on separate threads were used. Each thread contained 10 replicas with $\beta$ values multiplicatively distributed within the selected range. Each replica completed 4 iterations between $\beta$ swaps. For these swaps, replicas were swept from lowest to highest $\beta$; if the next highest $\beta$ replica had a higher or equal energy, then their two configurations were deterministically swapped. As such, if the lowest $\beta$ replica held the configuration with the lowest energy, then it could be swapped repeatedly to the highest $\beta$ replica. If 100 swap periods, 400 iterations, passed without the highest $\beta$ replica swapping, it swapped with the lowest $\beta$ replica.

The data in Fig. 3 represents results for the tai256c problem. This problem is unique in that the flow between any two of first 92 facilities is 1, and 0 between all other pairs. Thus, a p-dit whose dimension is in the first 92 will produce the same energy differential when swapped with any which are not. Similarly, a swap between two p-dits which both have a facility outside of the first 92, will always produce an energy differential of 0. These symmetries are used to simplify the work required for update calculations. Additionally, the automatic $\beta$ selection algorithm was not used.

## GPU-PIM Algorithm

Implementation details were generally similar to the CPU-PIM; however, 32 independent sets of replicas were used rather than 20. Once the $\beta$ selection algorithm was complete on CPU, these parameters were passed to persistent kernels running on the GPU, with the CPU only in charge of file I/O and terminating the program.

## GPU-CUPIM Algorithm

Implementation details were generally similar to the GPU-PIM, with 32 independent sets of replicas. As with the other implementations, each replica completed 4 iterations between $\beta$ swaps. For these swaps, replicas were swept from lowest to highest $\beta$; if the next highest $\beta$ replica had a higher or equal energy, then their two configurations were swapped. As such, if the lowest $\beta$ replica held the configuration with the lowest energy, then it could be swapped repeatedly to the highest $\beta$ replica. The same 100 swap-period timeout was used as with the CPU-PIM and GPU-PIM. Computationally, each replica used between $O(N)$ (for swapping operations) and $O(N^2)$ (for force and energy matrix updates) independent CUDA threads based on the phase of the algorithm.

During each iteration, as described using pseudocode in Supplementary Algorithm 1, the traditional PIM implementation calculates $\Delta E$ values for $N^2$ possible swap moves (swapping between any two p-dits is considered twice during each iteration) in groups of $N$ using incrementally updated $I$ values. In contrast, as described using pseudocode in Supplementary Algorithm 2, the GPU-PIM calculates all $\Delta E$ values for considered swaps at the beginning as a "snapshot" to approximate the energy change of sequential swaps, while removing the actual sequential dependency. This consequently creates the new possibility of collisions if two facilities wish to swap to the same position, requiring the addition of conflict resolution logic. Conflicts are resolved by choosing the more energetically favourable swap, or in the case of a tie, favouring the lower index p-dit which wishes to swap. Losers of collision resolution do not change position. The algorithms otherwise remain the same.

## Gurobi Baseline

Gurobi 12.0.2 was used as the comparison baseline for the results reported in Fig. 2. As with the PIM, 20 threads were used. The "SoftMemLimit" and "NodefileStart" parameters were set to 40 GB. The "BestObjStop" parameter was set to 0.001 above the best reported solution in the QAPLIB. A $n \times n$ binary matrix was used, with a value of 1 at position $(i, j)$ indicating that facility $i$ was assigned to location $j$ and 0 indicating otherwise. A total of $2n$ constraints were added to ensure each row and column sum was equal to $1$. The non-zero products of each pair of elements in the $F$ and $D$ matrix, multiplied by the two corresponding variables from the binary matrix, were added as terms in a quadratic objective function.

While Gurobi 12.0.2 was still used for the results of Fig. 3, a different formulation, based on previously reported results was used[11]. This formulation takes advantage of the symmetry of the problem. Like with the previous results, $20$ threads were used, and the "SoftMemLimit" and "NodefileStart" parameters were set to $40$ GB. Additionally, the "MIPFocus", and "PreQLinearize" parameters were set to $3$ and $0$ respectively. A binary vector of length $n$ was used to represent the state of the problem. A constraint that all elements of this vector sum to $92$ was used. The distance between each pair of locations, multiplied by the corresponding entries in the binary state vector, were used as terms for the quadratic objective function.

## Hardware and Data Reporting Details

All reported results were run on a machine with an AMD Ryzen 9 7900X 12-Core Hyperthreaded Processor, 64 GB of DDR5 4800MT/s RAM, and a NVIDIA GeForce RTX 4070. For the CPU-PIM implementation and the GPU implementations, setup time for the main portion of the algorithm was not included in the reported runtime, but the automatic $\beta$-selection procedure was. For Gurobi, the problem setup time was not included in the reported runtime, but the presolve time was. For all, runtime was counted as when the algorithm first logged the best result it found. For the PIMs, this is reported to the nearest millisecond. However, for Gurobi, this is reported to the nearest second. This difference exaggerates the difference between the PIM and Gurobi in Fig. 2c for instances where Gurobi performs better. The PIM reports solution quality with full precision. Gurobi, on the other hand, reports only the final solution quality with full precision, and reports intermediate solutions with variable precision. In cases for which this produced uncertainty for when the final solution was discovered, the earliest possible timestamp was used. As a further consequence of this reporting, the values for Gurobi in Fig. 3 are rounded, typically to the nearest 1,000, except for the final value.

For Fig. 2, all QAPLIB instances for which solutions were provided, except tai256c, were considered. After running the sweep, it was discovered that the solution file provided for kra32 did not match the true optimal value, so this data was manually corrected.

The results for each solver on all the QAPLIB instances considered, excluding tai256c, is shown in Supplementary Table 1. Included are results for the Hexaly[39] optimizer, however a memory limit was not able to be set for it, and no time data was reported.

## Acknowledgements

The work at Northwestern University was supported by the U.S. National Science Foundation under awards 2322572, 2311296, 2433807, and 2425538.

## Author Contributions

C.D. and A.G. conceived the idea of the project. C.D. formulated the problem and designed the methodology. C.D. was responsible for making the CPU implementation code. C.D. and C.M.B. were responsible for making the GPU implementation code. J.A. assisted with the data presentation. G.F., E.W., and P.K.A. provided support in the experimental design and data analysis. The manuscript was written by C.D. with contributions and feedback from all other authors. The study was performed under the leadership of P.K.A..